%% file: main.tex
\documentclass{article}

\input{config}

\begin{document}

\title{Applying separative non-negative matrix factorization to extra-financial data.}

\author[$\dagger\circ$]{P. Fogel}
\author[$\dagger$]{C. Geissler}
\author[$\dagger$]{P. Cotte}
\author[$\star$]{G. Luta}

\affil[$\dagger$]{Advestis, 69 Boulevard Haussmann, 75008 Paris, France}
\affil[$\dagger$]{\textit{\{cgeissler, pcotte\}@advestis.com}}
\affil[$\circ$]{\textit{paul\_fogel@hotmail.com}}
\affil[$\star$]{Georgetown University, 3700 O St NW, Washington, DC 20057, USA}
\affil[$\star$]{\textit{george.luta@georgetown.edu}}
\affil{\small{The first two authors contributed equally to the article.}}

\maketitle

\begin{abstract}
    We present here an original application of the non-negative matrix factorization (NMF) method, for the case of extra-financial data. These data are subject to high correlations between co-variables, as well as between observations. 
    NMF provides a much more relevant clustering of co-variables and observations than a simple principal component analysis (PCA). In addition, we show that an initial data separation step before applying NMF further improves the quality of the clustering.
\end{abstract}


\section{Introduction}\label{sec::introduction}

    A question frequently encountered in economics is that of business classification. The financial industry (in particular index providers) has developed several standards for classifying companies based on their dominant industrial activity, such as the Global Industry Classification Sector (GICS \textregistered). These classifications are widely used by investment fund managers as a tool for diversifying economic risk, or alternatively to create thematic portfolios focused on one or a small number of industry sectors.\\
    
    The purpose of this paper is to outline a method for classifying companies based on their extra-financial profiles. To this end, we will use scores from extra-financial rating agencies. These scores quantify the behaviour of issuers in the three main pillars of Environment (E), Social (S) and Governance (G).\\ 
    
    For the data-scientist, these data constitute a rich and complex field of study. The richness is linked to the variety of themes noted, ranging from the treatment of employees and suppliers, to the carbon footprint and the independence of the Board of Directors. These ratings provide information that cannot be derived from the industrial sector alone, and provide a new viewpoint to discriminate between companies. \\
    
    The complexity stems from the fact that those data are not provided uniformly for all issuers, as some specific scores only make sense for very specific sectors. On the other hand, behavioural scores are both numerous (typically between 50 and 100) and sometimes highly correlated.  \\
    
    We describe here an unsupervised clustering experiment based on a matrix of extra-financial data. The data constituting the matrix are normalized between 0 and 100, and therefore lend themselves well to the use of non-negative matrix factorization (NMF) methods. This factorization provides a grouping of both co-variables and companies. The results are compared with those of a Principal Component Analysis (PCA). We also show, following a path initiated by \cite{IJERPH2016}, that a tensor variant of NMF can be used to isolate groups of issuers characterized by either high or low scores on some co-variables, while ordinary NMF would mainly focus on groups of high scores.

\section{Presentation of the data}\label{sec::data_presentation}

    We use ESG scores from the research of Vigeo-Eiris, whom we thank here for making these data available for this study.\\
    
    The data consist of 77 scores spread over the three pillars E, S and G. They concern 699 European stocks. The scores are positive numbers between 0 and 100, with 0 representing the worst possible score and 100 the best. These quantities do not have absolute significance but are essentially used to establish a ranking between issuers.

    \subsection{Nomenclature of variables}\label{ssec::nomenclature}

        We will denote by $\X^{ESG}$ the matrix formed by the 77 variables or 'features' as columns. This matrix is of type $(699, 77)$. The generic term $\X^{ESG}_{i,j}$ is the value of the feature $j$ for stock $i$.\\
        
        Vigeo’s behavioural scores are grouped into 6 areas, with taxonomy being reflected in the denomination of variables:
        \begin{itemize}
            \item Environment: variables names starting with \textbf{env\_}.
            \item Corporate Governance: "\textbf{cg\_}"
            \item Social: Business Behaviour: "\textbf{c\_s\_}"
            \item Social: Human rights: "\textbf{hrts\_}"
            \item Social: Human Resources: "\textbf{hr\_}"
            \item Social: Community involvement: "\textbf{cin\_}"
        \end{itemize}
        More information about the scoring methodology can be found at: 
        \url{https://vigeo-eiris.com/about-us-2/methodology-quality-assurance/}.\\
        
        The interesting feature here is that we have a prior fundamental knowledge about variables that will be used as a benchmark to evaluate the relevance of any NMF-based clustering. Similarly, the fundamental information about the industrial sector the companies belong to will also be used as a prior knowledge.

    \subsection{Missing data}\label{ssec::missing_data}

        Variables are scores that rate the company's activity according to different criteria. Some of these criteria have no relevance outside a particular industry sector, and are therefore not reported for all issuers.\\
        
        Table \ref{table:FillBySector} illustrates the fill rate (number of stocks for which the variable is filled in, divided by the number of stocks). The fill rate is calculated as an average per industry sector. \\
        
        \begin{table}[ht]
            \centering
            \begin{tabular}{|l|c|}
                \hline
                & \textbf{Average fill rate} \\ \hline
                Consumer Staples & 88.41\,\% \\ \hline
                Consumer Discretionary & 85.52\,\% \\ \hline
                Materials & 84.36\,\% \\ \hline
                Health Care & 83.89\,\% \\ \hline
                Utilities & 83.54\,\% \\ \hline
                Industrials & 82.37\,\% \\ \hline
                Energy & 79.22\,\% \\ \hline
                Real Estate & 79.02\,\% \\ \hline
                Communication Services & 78.03\,\% \\ \hline
                Information Technology & 77.18\,\%\\ \hline
                Financials & 76.21\,\% \\ \hline
                \end{tabular}
            \caption{\label{table:FillBySector} Sectors sorted by decreasing average fill rate, the average being taken over all features. The global average fill rate is 81.1\,\%}
        \end{table}
        
\begin{sidewaysfigure}[htp!]
            \centering
            \includegraphics[width=0.9\textwidth]{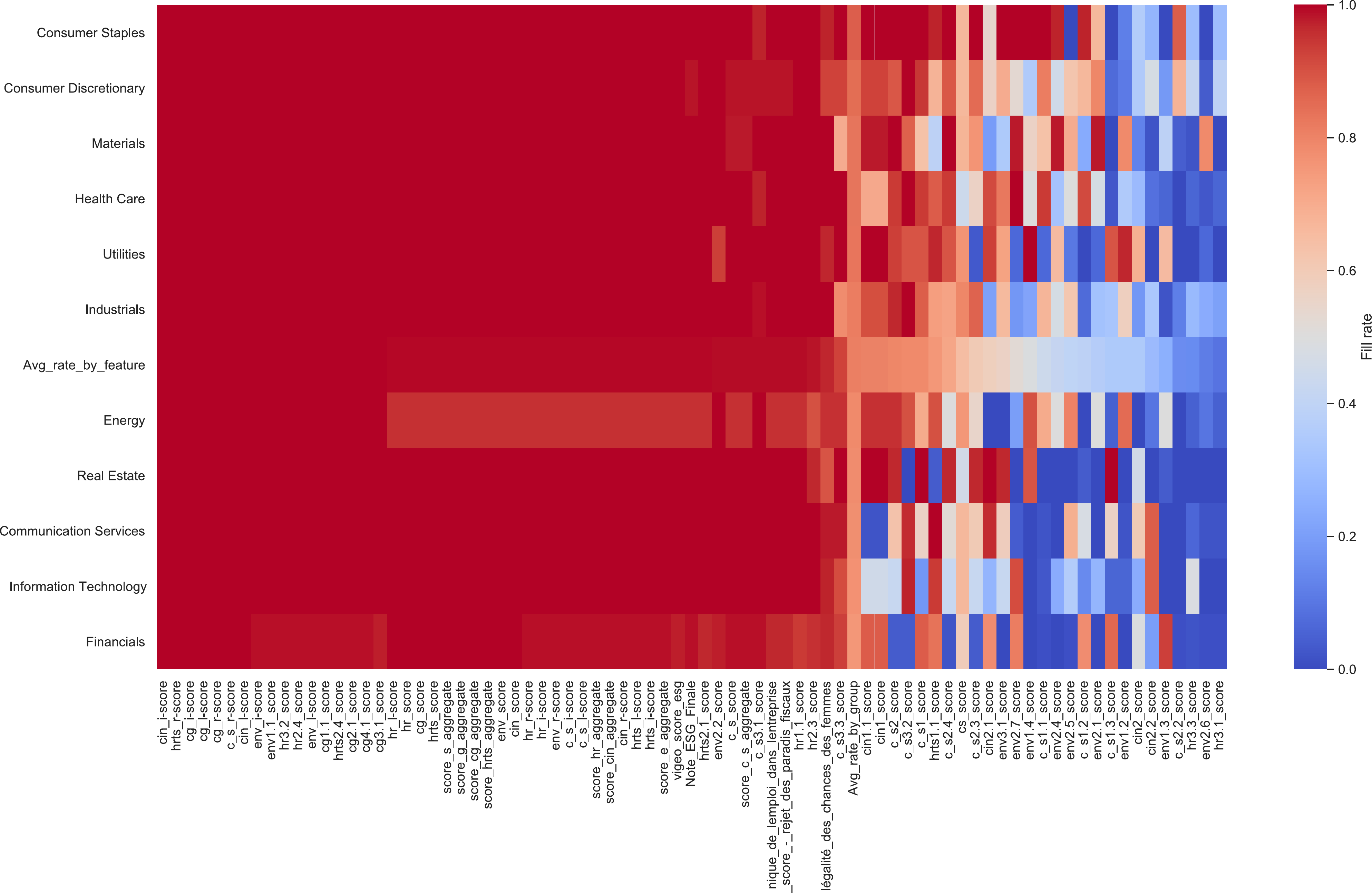}
            \caption{}
        \caption{\label{fig:Fill rate by sector}Each cell represents the average fill rate of a given feature within a given sector.}
        \end{sidewaysfigure}

\begin{figure}[htp!]
        	\centering
        	\includegraphics[width=0.8\textwidth]{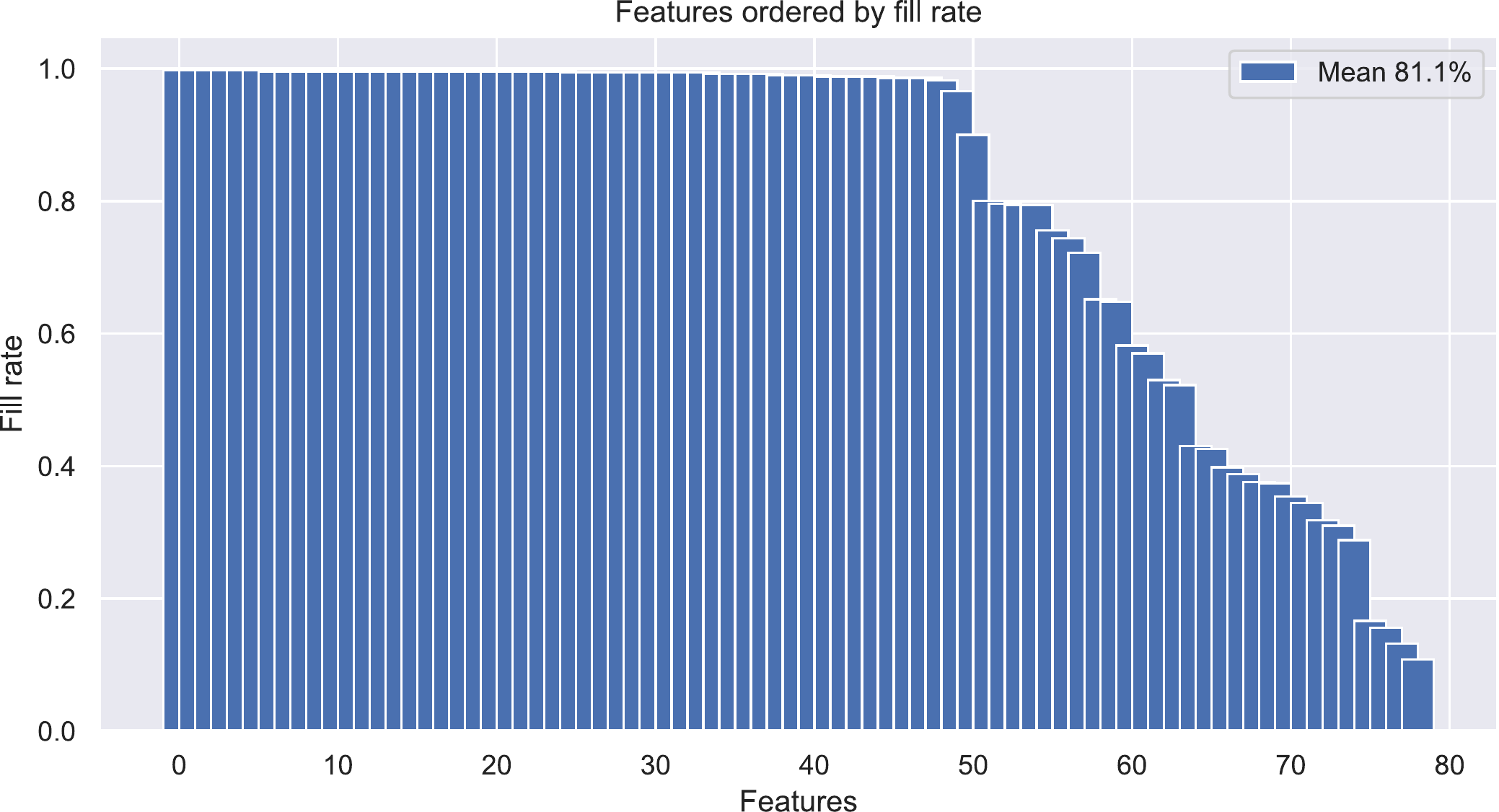}
        	\caption{\label{fig:Fill rate distribution}The average fill rate by feature.}
        \end{figure}

        The next two charts give a visual summary of the average fill rates grouped by sector (Figure \ref{fig:Fill rate by sector}) and by features (Figure \ref{fig:Fill rate distribution}). The sectors are presented along lines, by decreasing rate, while the variables are presented in columns, also by decreasing rate. As shown on chart \ref{fig:Fill rate by sector}, the consumer staples sector has the least amount of missing data. In contrast, several environment-related scores in the finance or telecommunications sector have little or no information at all, i.e. a fill rate close to zero. twenty two features out of 77 have a fill rate less than 75\,\%, the average being 81.1\,\%.
        
        
        \FloatBarrier
        
        Before performing any statistical analyses such as linear regressions, it is necessary to apply a missing values imputation strategy. A default strategy is to fill in the missing data for a variable with the mean or median, taken over the entire sample, of the non-missing values of that variable. 

\section{A first analysis using Principal Components Analysis (PCA)}\label{sec::PCA}

    Principal component analysis of a $\X$ data matrix focuses on the uncorrelated directions that explain the highest proportion of variance in the data. The directions are linear combinations of the original variables, which are mutually orthogonal. These principal directions are none other than the eigenvectors of the $\M = \X\X\tran $ matrix.\\
    
    In the case of the $\X^{ESG}$ matrix, we need to fill in the missing data with non-empty numerical values, because PCA algorithms are not meant to handle missing values. To do this, we apply the most neutral imputation strategy with respect to the objective of explaining variance: for each variable, a missing observation is replaced by the average of the variable, calculated on the set of non-missing data. \\
    
\begin{figure}[htp!]
    	\centering
    	\includegraphics[width=1.1\textwidth]{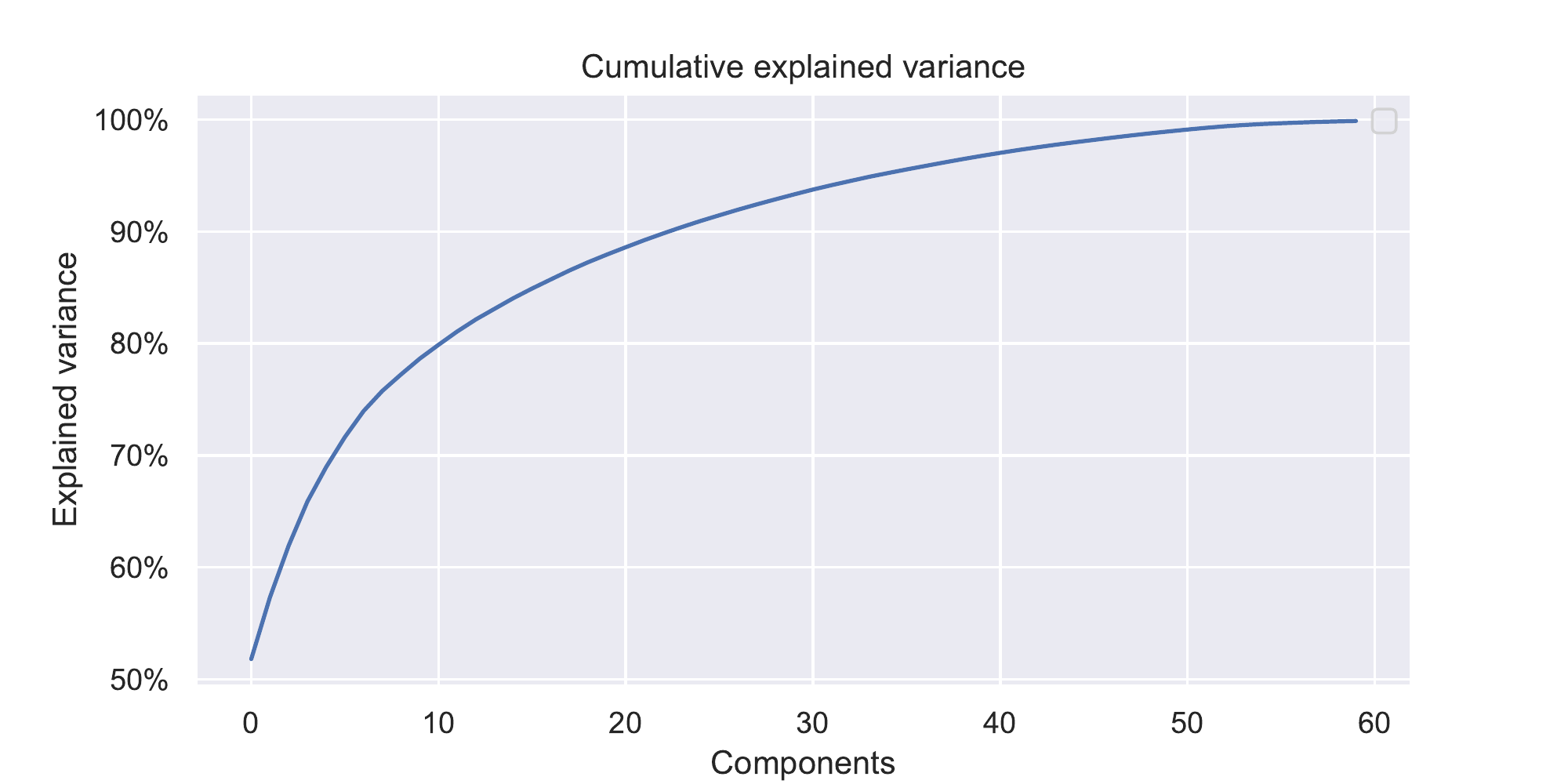}
    	\caption{\label{fig::pca explained variance} Cumulative explained variance of the first n components or eigenvectors.}
    \end{figure}

\begin{figure}[htp!]
    	\centering
        \includegraphics[width=1.1\textwidth]{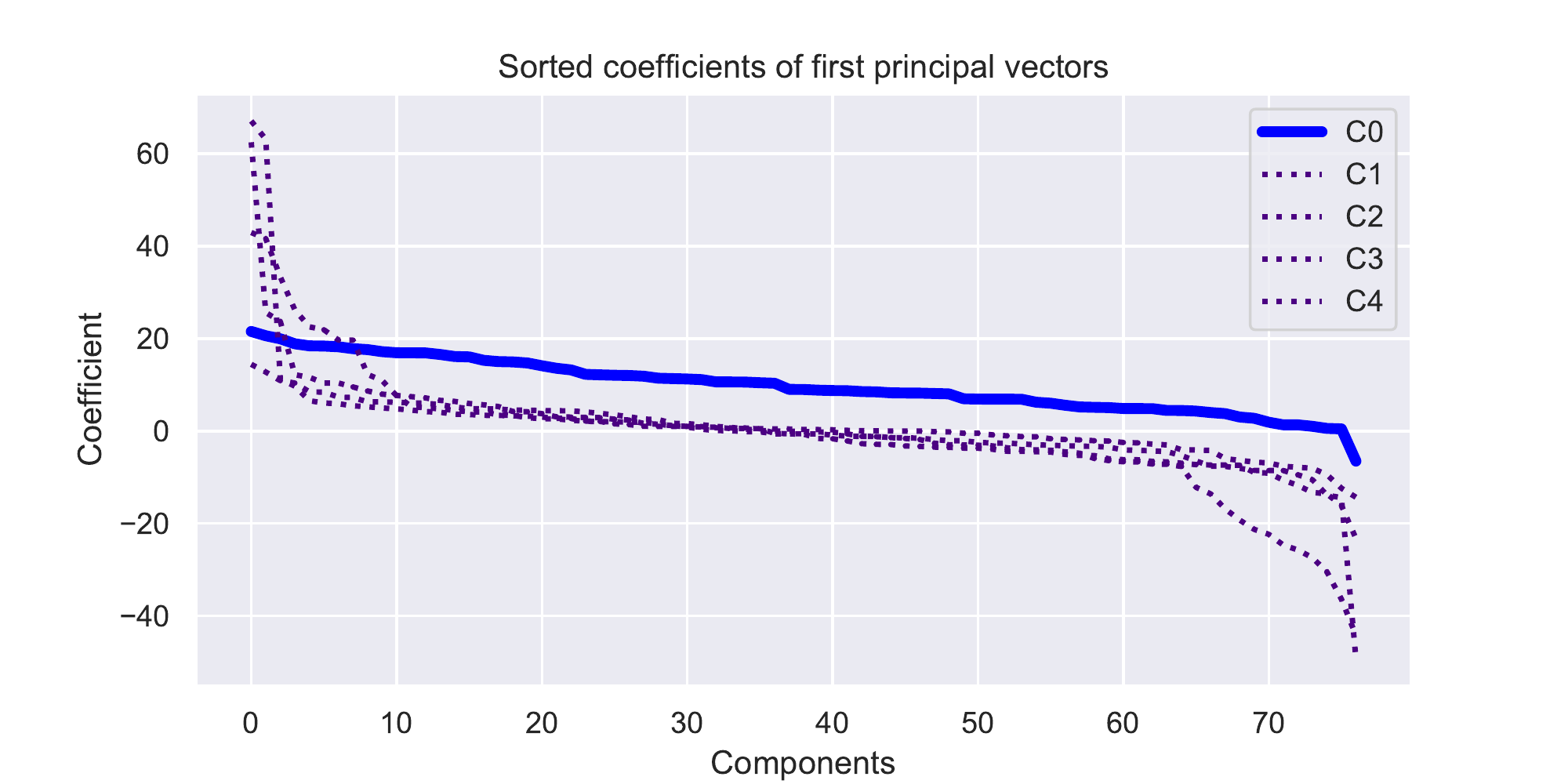}
    	\caption{\label{fig::pca sorted coeffs} Coefficients of the first 5 components, sorted by decreasing order.}
    \end{figure}

    A first look at the results of a standard PCA allows the following observations to be made:
    \begin{itemize}
        \item The observation of the cumulated variance explained according to the rank of the component shows that it increases rather slowly. At least twenty two components are needed to explain 90\,\% of the total variance (Figure \ref{fig::pca explained variance}).
        \item The first component has coefficients that are all positive except one. All other components, which are mutually orthogonal, have positive and negative coefficients, at least half of which are greater in absolute value than 10 \% of the largest coefficient. (Figure \ref{fig::pca sorted coeffs}).
    \end{itemize}
    
    \FloatBarrier
    
    Except for the first component (thick blue line) for which all coefficients but one are positive, all other components have coefficients of both signs, which makes their interpretability difficult. 
    \FloatBarrier

\section{Factoring the data through Non negative Matrix Factorization (NMF)}\label{sec::NMF}

    \subsection{Principles}\label{ssec::NMF_principles}

        Let us assume the available data are represented by an $X$ matrix of type $n \times f$, i.e. $n$ rows and $f$ columns. We also assume that these elements are positive or null and bounded. With that in mind, we can re-write $X$ (here shown as a $3\times4$ example matrix, with rows and columns indices starting at 0) as
        \begin{equation}\label{eq:example_nmf}
            \underbrace{\begin{pmatrix}
                x_0^0 & x_0^1 & x_0^2 & x_0^3 \\
                x_1^0 & x_1^1 & x_1^2 & x_1^3 \\
                x_2^0 & x_2^1 & x_2^2 & x_2^3 \\
            \end{pmatrix}}_{X} = 
            \underbrace{\begin{pmatrix}
                w_0^0 & w_0^1 \\
                w_1^0 & w_1^1 \\
                w_2^0 & w_2^1 \\
            \end{pmatrix}}_{W}
            \underbrace{\begin{pmatrix}
                h_0^0 & h_1^0 & h_2^0 & h_3^0 \\
                h_0^1 & h_1^1 & h_2^1 & h_3^1 \\
            \end{pmatrix}}_{H\tran} + 
            \underbrace{\begin{pmatrix}
                \epsilon_0^0 &\epsilon_0^1 & \epsilon_0^2 & \epsilon_0^3 \\
                \epsilon_1^0 &\epsilon_1^1 & \epsilon_1^2 & \epsilon_1^3 \\
                \epsilon_2^0 &\epsilon_2^1 & \epsilon_2^2 & \epsilon_2^3 \\
            \end{pmatrix}}_{\epsilon=X - W H\tran}
        \end{equation}
    A non-negative factorization of $X$ (\cite{LeeSeung1999}) then solves the following problem:\\
        Find two matrices $W$ of type $n\times c$ and $H$ of type $f \times c$ minimizing the quadratic norm of the last term of the right hand of Equation \eqref{eq:example_nmf}
        \begin{equation}
            ||\epsilon||^2_2 \coloneqq \norm{X - W H\tran}^2_2
            \label{eq::objective function}
        \end{equation}
        $H$ and $W$ being subject to the constraints of only having non-negative entries.
        
        A few remarks:
        \begin{itemize}
            \item The quadratic norm used here is restricted to numeric entries (i.e. represents the square root of the sum of squared non-missing values). This slight extension of the definition allows to deal with input matrices having some missing values.
            \item The objective function in \ref{eq::objective function} may also include additional by L1 and L2 penalization terms applied to the W and H matrices (\cite{Hoyer2004}). The aim of this penalization is to eliminate noise coming from small and generally unstable coefficients, and to concentrate the signal on the most significant contributors. This objective is called \textbf{sparsity}.
            \item This decomposition is an approximation, not an equality. Any matrices $W$ and $H$  that are solutions to \eqref{eq::objective function} minimize the quadratic error of the approximation.
            \item It is common in the literature to refer to rows of $W$ matrix as \textbf{scores}, and to those of $H$ as \textbf{loadings}.
            \item The decomposition is not necessarily unique, even with the positivity constraint. One can indeed use any $D$ matrix having non-negative coefficients, as well as its inverse $D^{-1}$, and then obtain identical decompositions of the form:    $X = W D D^{-1}H\tran + \epsilon$. Different examples of such a transformation would be: 
            \begin{itemize}
                \item A change of unit in the variables, with a diagonal D matrix representing the equivalence coefficient between two measuring systems.
                \item A permutation in the components, D being then represented by a permutation matrix.
            \end{itemize}
            \item $c \leq \min(n,f)$ is an integer representing the number of selected components, also called the \textbf{rank} of the decomposition. Like in most clustering methods, this parameter represents a degree of freedom; determining the most adequate number of components requires an additional optimality criterion than the one for error, as the error generally decreases with the rank.
            \item The $n$ rows of $X$ will be referred to as \textbf{observations}, the $f$ columns of $X$ as \textbf{variables} or \textbf{features}, and the $c$ columns of $W$ as \textbf{meta-variables} or \textbf{meta-features}.
        \end{itemize}
        
        It is customary to consider the rows of $X$ as $n$ observations of objects from the same category (e.g. customers, patients, companies, etc). Each object has $f$ attributes, and the data contained in the matrix correspond to the values taken by the attributes for the corresponding objects.\\
        
        The main effect of this decomposition is to decrease the information necessary to describe an observation. The original observations of the $X$ matrix can be recovered, at the cost of an approximation, as linear combinations with positive coefficients of the rows of the $H$ matrix. A reduction in size has thus been achieved: part of the information contained in the original matrix, comprising $f$ features per individual, can be approximately summarized by a smaller set of $c$ features per individual.

    \subsection{A first NMF run}\label{ssec::first_trial}

        We start with a first NMF decomposition on the $500\times77$ scores matrix $X$. Two prior steps are handling missing values in $X$, and choosing a rank $c$. \\
        
        We use the Python package \textbf{nmtf} developed by one of the authors. This package is public and available at \url{https://pypi.org/project/nmtf/}. It can be directly installed with the pip command: \textit{pip install nmtf}.\\
        
        The optimisation algorithm used in this package takes advantage of the extended definition of the quadratic norm and can handle missing values in input matrices, as opposed to the NMF version 0.17 of Scikit-Learn, which raises an error in case of missing entries in the matrix.
        
        This ability to handle missing values avoids having to deal with the delicate issue of imputation of these missing values. In the field of extra-financial scores, a missing value is not simply data for which a measurement is missing because it did not take place. It can be data whose measurement simply does not make sense. To give a concrete example from the environmental domain, scores for waste reprocessing or contribution to deforestation are not relevant for some industrial sectors.\\
        
        Another difference between nmtf and Scikit-Learn lies in the way sparsity is specified. In Scikit-Learn, the objective function is modified by addition of penalty coefficients that have an indirect effect on the sparsity of the W and H matrices. In nmtf, on the contrary, the sparsity objective on H or on W is provided as a parameter to the model, as a positive coefficient less than 1. For example, a parameter value of 0.9 on H will provide, if the resolution is possible, an H matrix with about 90\% of zero entries.\\ 

        The rank is set here at 6. This choice is not totally arbitrary as a first trial, it is exactly the number of topics set by Vigeo-Eiris for the taxonomy of the scores.\\
        
        The solution $\hat{X}$ provided by the algorithm satisfies:
        $\norm{X - W H\tran}^2_2 = 5.7\times 10^{-2} \norm{X}^2_2$.\\
        
        In other words, the squared approximation error for rank $c = 6$ is 5.7\,\% of the squared norm of $X$. Referring to the performances of PCA on chart \ref{fig::pca explained variance}, it is necessary to aggregate the first 32 factors in order to explain a fraction of $100\,\% - 5.7\,\% = 94.3\,\%$ of the total variance. This illustrates one of the main strengths of NMF: relaxing the requirement of factor orthogonality provides in return a much stronger explanatory power.

    \subsection{Clustering observations and features}\label{ssec::clustering}

        An interesting induced property of NMF is the ability to cluster variables and observations in a natural way.\\
        
        Regarding the clustering of features, the principle is to link each of the $f$ variables with the meta-variable (indexed between 1 and $c$) having the highest coefficient on this variable. Similarly, each observation is assigned the index of the meta-variable with the highest contribution to the observation.\\
        
        This grouping scheme has the disadvantage of being dependent on the particular system of units retained for the meta-variables, i.e. a transformation by an arbitrary $D$ matrix.
        In order to get around this disadvantage, the authors of [5] introduce the notion of leverage, which represents "the ability of a row or column to specifically exert an influence on a particular element without a corresponding increase in the influence on other components". Leverage is a dimensionless quantity between 0 and 1. The clustering scheme based on maximum leverage is more stable than that based simply on the maximum coefficient.\\
        
        The most visible benefit that we obtain from the non-negative matrix is the replacement of the initial 77 features with a reduced subset of 6 meta-features; the initial features can themselves be written as positive linear combinations of these meta-features.\\
        
        \begin{figure}[htp!]
        	\centering
            \includegraphics[width=1.1\textwidth]{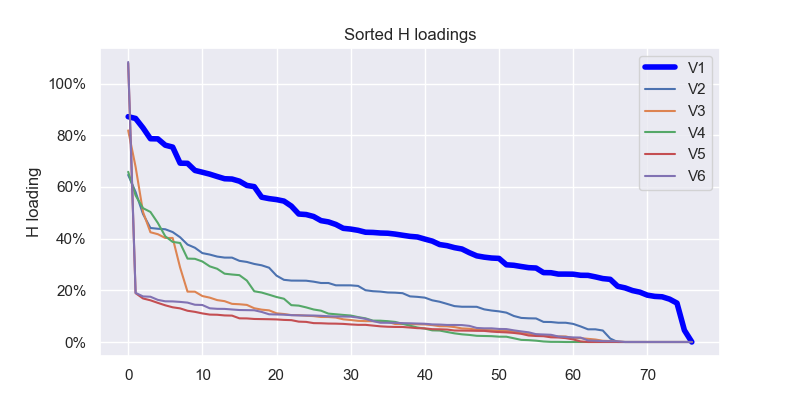}
        	\caption{\label{fig::sorted_h_loadings} Coefficients of the 6 meta-features sorted by decreasing order.}
        \end{figure}
        
        Figure \ref{fig::sorted_h_loadings} shows the distribution of the coefficients (the ‘loadings’) of the initial 77 features over the 6 new meta-features, sorted by decreasing order. We observe that as the rank of the meta-features increases, fewer and fewer of the original features have significant coefficients on them.\\



\section{Introducing Separative NMF}\label{sec::separative_NMF}

    One of the few weaknesses of NMF is its asymmetry of treatment between high and low values. By construction, standard NMF will group observations by looking for the meta-features having highest scores. The attribution of a cluster will be more discriminant for companies characterized by a high score on a meta-feature (e.g. high environmental scores), and less discriminant for companies characterized by a low score. But when it comes to ESG scores, knowing that a company has a low score on a meta-feature is just as important as the information provided by a high score. \\
    
    In \cite{IJERPH2016}, an interesting variant of NMF is proposed in order to go around this obstacle. The method is called PosNegNMF. The idea is to separate the original matrix and to rewrite it as the sum of three components:\\
    \begin{equation}\label{eq::SNMF}
        X = X^+  - X^- + X_0
    \end{equation}
    with the following notations:
    \begin{itemize}
        \item $X_0$ is the "baseline", or the matrix of columns-wise medians of X.
        \item $X^+ = Max(0, X - X_0)$, the matrix of positive high deviations from $X_0$.
        \item $X^- = Max(0, X_0 - X)$, the matrix of positive low deviations from $X_0$.
    \end{itemize}
    Having done this, PosNegNMF will perform a single decomposition of the matrix containing positive and negative deviations. This is equivalent to two independent decompositions, yielding two H matrices with a common W matrix. More precisely:
    \begin{equation}\label{eq::PosNegNMF}
        \mathds{X} =_{def} [X^+ | X^-] = W [H^+ | H^-]\tran + \epsilon
     \end{equation}
    \begin{equation}
        X = X^+  - X^- + X_0 = W (H^+ - H^-)\tran + X_0 + \epsilon
    \end{equation}.
    The notation $[A\space|\space B]$ represents the $r \times 2c$ horizontal concatenation of two $r \times c$ matrices A and B. The first equation describes the NMF decomposition of $\mathds{X}$.
    The second one reconstructs the original matrix as the sum of baseline, high and low deviations.\\
    
    The new approach proposed here relies on the more general $k(=3)-$dimensional algorithmic version of NMF, known as Non-negative Tensor Factorization (NTF, \cite{Cichocki09}), to approximate the original matrix by a three-dimensional $n \times f \times 2$ tensor $\mathds{X}$:
    \begin{equation}\label{eq::NTF}
        \mathds{X} =_{def} [X^+ | X^-] = W \otimes H \otimes Q + \epsilon
    \end{equation}

    Note that all elements of $\mathds{X}$ are non-negative. Similarly to the NMF-based algorithm described above, this decomposition provides a canonical clustering based on the meta-features provided by $H$ matrix. Let us emphasize that this $H$ matrix is common to the two slices of the tensor $\mathds{X}$. In other words, what we obtain is not the same as PosNegNMF, which performs two NMF decompositions on $X^+$ and $X^-$ respectively. The $H$ matrix describes common features having an influence on either higher or lower deviations of the scores from their median. \\
    
     PosNegNMF yields an approximation $X^+ - X^- = W H\tran $ with $H = H^+ - H^-$ . Thus, each column of H may have signed entries, which complicates interpretation since the features under study are non-negative scores. In contrast, our NTF-based approximation yields a non-negative H matrix in which each column describes common features that have an impact on either higher or lower deviations of the scores from their median, as determined by the sign of the difference $Q[1, \text{column}] - Q[2, \text{column}]$. The non-negativity of the H-loadings makes interpretation as straightforward as it is for the standard NMF. For this reason, we propose to call our approach Separative NMF, or S$^{2}$NMF, since it aims at separating patterns of features with either positive or negative deviations.
    
    \subsection{Precision and sparsity}\label{ssec::precision_and_sparsity}
        
        We will now evaluate the comparative advantage brought by S$^{2}$NMF, by varying the level of control of parsimony of the variables. More precisely, we observe how the approximation error and the composition of the meta-features evolve as the model is constrained to work with the most significant features. In order to focus on the sole effect of the parsimony, we start from a fully populated data matrix. Missing data are replaced for each feature, by the average of non-missing values of this feature. Doing this prevents from any unwanted numerical side-effect that can appear in the presence of a sparsity constraint applied to a matrix with empty entries.\\
        
        To perform this experiment, we proceed as follows:
        \begin{itemize}
            \item The number of components is set to 6.
            \item We run in parallel an ordinary NMF model with its S$^{2}$NMF counterpart.
            \item Both models are first launched with a sparsity target of 0.9 on the matrix H of meta-features. To achieve such a sparse representation, each vector of H is approximated after each update of the numerical resolution by a close vector (in the sense of quadratic norm) with the required degree of sparsity (\cite{Potluru2013}, \cite{Hoyer2004}). By doing so, some features show zero coefficients on all meta-features in matrix H. This is the \textbf{sparse} version of the model.
            \item The features showing only null loadings at the previous step are dropped from the matrix X, and the model is run again on the reduced matrix with no sparsity constraint: this is the \textbf{restricted version} of the model.
        \end{itemize}
        
\begin{figure}
        	\centering
            \includegraphics[width=\textwidth]{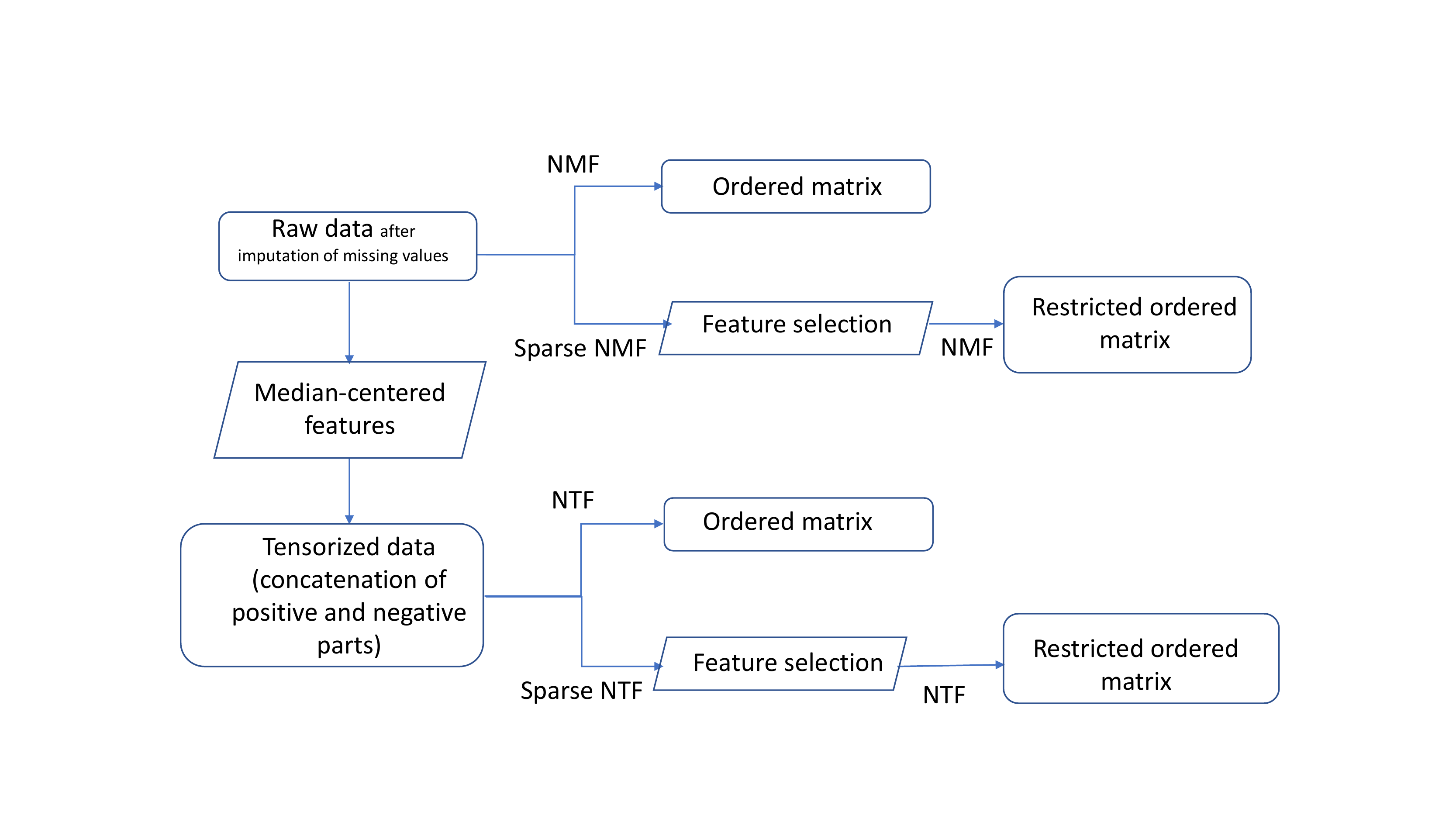}
        	\caption{\label{fig::process_scheme} A schematic representation of the various factorizations applied to the initial data. Rectangles with rounded corners represent input or output data, while rectangles with slanted edges represent algorithmic treatments.}
        \end{figure}

        Figure \ref{fig::process_scheme} summarizes the steps of this process.
        
        Table \ref{table::nmf_vs_snmf_error_table} presents the results of this experiment. The columns of the tables are defined as follows:
        \begin{itemize}
            \item Dimension is the number c of components, here set at 6 for all models.
            \item Features refers to the number of features effectively included in the factorization. 
            \item $H$-sparsity is a tuning parameter of the NMF/S$^{2}$NMF algorithm, representing a target level of sparsity, or a proportion of null values, in matrix $H$.
            \item The Error column is the relative squared error of the approximation.
        \end{itemize}

        \begin{table}[ht!]
            \centering
            \begin{tabular}{|c|c|c|c|c|}
                \hline
                \textbf{Model} & \textbf{Dimension} & \textbf{Features} & \textbf{$H$-Sparsity} & \textbf{Error} \\ \hline
                \textbf{Full NMF}  & \textbf{6} & \textbf{77} & \textbf{0} & \textbf{3.54 \%} \\ \hline
                \textbf{Sparse NMF}  & \textbf{6} & \textbf{77} & \textbf{0.9} & \textbf{155.2 \%} \\ \hline
                \textbf{Restricted NMF}   & \textbf{6} & \textbf{36} & \textbf{0}   & \textbf{2.1\%}   \\ \hline
                \textbf{Full S$^{2}$NMF} & \textbf{6} & \textbf{77} & \textbf{0.0} & \textbf{4.41 \%}    \\ \hline
                \textbf{Sparse S$^{2}$NMF} & \textbf{6} & \textbf{77} & \textbf{0.9} & \textbf{9.2\%}    \\ \hline
                \textbf{Restricted S$^{2}$NMF}  & \textbf{6} & \textbf{32} & \textbf{0}   & \textbf{4.8 \%}   \\ \hline
            \end{tabular}
            \caption{\label{table::nmf_vs_snmf_error_table} Comparison of NMF and S$^{2}$NMF in sparse and dense versions.}
        \end{table}
        
        We see that the ordinary NMF model collapses  when the sparsity is constrained at the high level of 90\%. The mean squared error of the factorization becomes greater than the variance of the original data. In order to recover an acceptable error level, we have to drop 41 features out of 77 that do not contribute to the composition of any meta-feature, leaving us with 36 significant features.
        By contrast, starting directly with a Separative NMF approach in sparse mode, effectively reduces the error to an acceptable level of 9\%. Forty five features are flagged as irrelevant, leaving the dense model with a remaining 32 columns. 

    \subsection{Clusterization efficiency}\label{ssec::efficiency}
        
\begin{figure}[!htb]
            \begin{minipage}{0.49\textwidth}
            	\centering
                \includegraphics[width=0.5\textwidth]{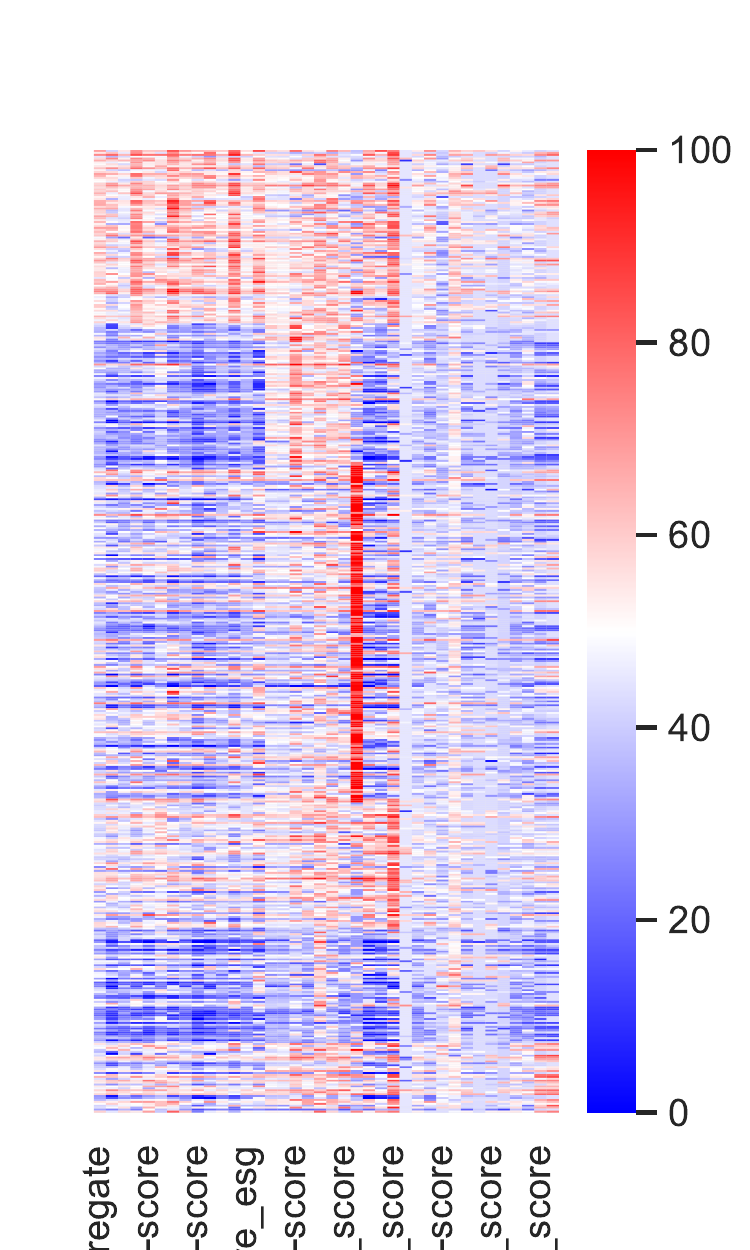}
            	\caption{\label{fig::restricted_nmf_heatmap} The original matrix reordered according to NMF clusters, and restricted to the 37 features preserved by the sparsity condition.}
            \end{minipage}
            \begin{minipage}{0.49\textwidth}
            	\centering
                \includegraphics[width=0.5\textwidth]{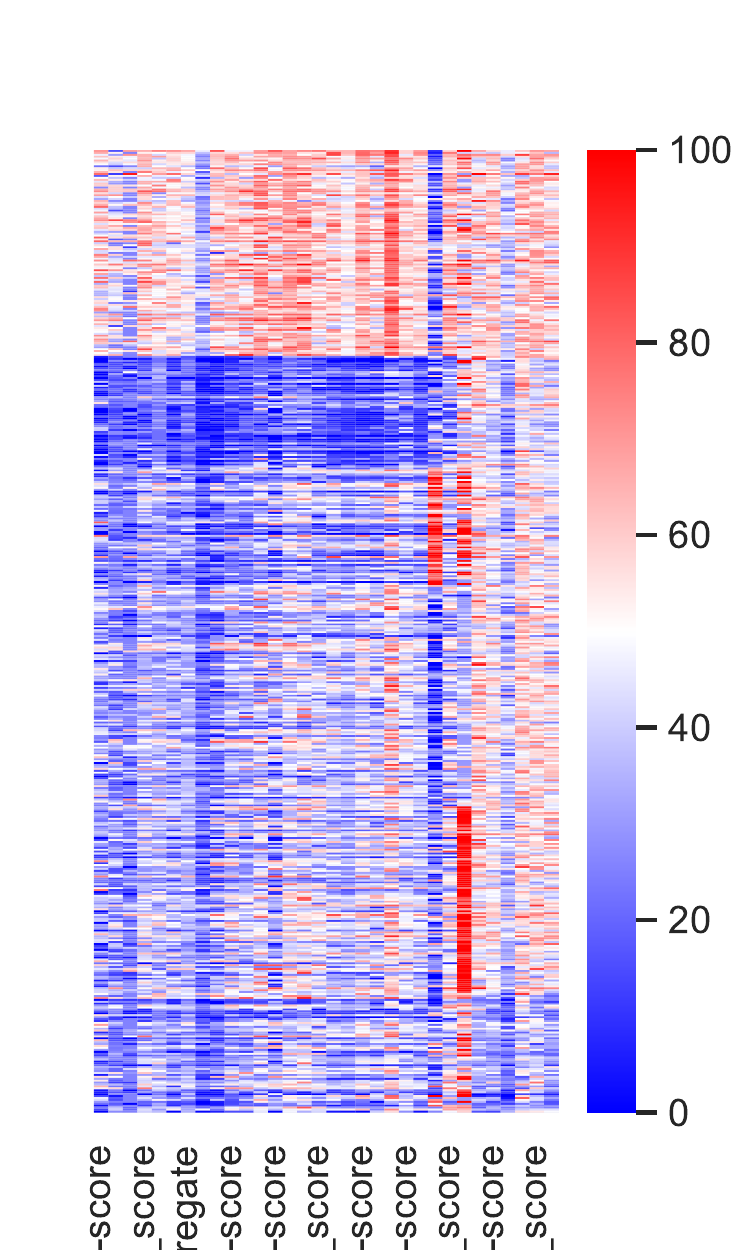}
            	\caption{\label{fig::restricted_SNMF_heatmap} The original matrix reordered according to S$^{2}$NMF clusters, and restricted to the 32 features preserved by the sparsity condition.}
            \end{minipage}
        \end{figure}

        The advantage of S$^{2}$NMF over NMF is more easily seen by looking at the ordered heatmaps presentations of the matrix.
        Figures \ref{fig::restricted_nmf_heatmap} and \ref{fig::restricted_SNMF_heatmap} show NMF- and S$^{2}$NMF- based heatmaps of the restricted matrix. The restriction takes out the features that have been eliminated in the first step with a target sparsity set at 90\,\%.
        
        
        It is visually clear that S$^{2}$NMF has a greater ability than NMF to discern and separate clusters, at both ends of the score range (blue or red aggregates).\\
        
        If we try to formalise this visual intuition, the mathematical translation is that the distribution of scores within clusters is more homogeneous than in the totality of the data. This homogeneity can be quantified thanks to the notion of entropy.
        
        Let us recall its definition: for any probability distribution $p$ on a finite set: $$X=\{x_1, \ldots, x_n\}$$ the entropy of the distribution is given by 
        \begin{equation}\label{eq::entropy}
            E(p) = \sum_{i=1}^{i=n}p(x_i)\times\log\left(\frac{1}{p(x_i)}\right)  
        \end{equation}
        Information theory tells us that $\log(1/p(x))$ is (up to an additive constant) the optimal number of bits required to encode the value $x$, given its probability of occurrence $p(x)$. From this viewpoint, $E(p)$ appears to be the expected encoding length of a random pick from $X$. The minimum possible entropy is 0, attained for a distribution concentrated at 100\,\% on a single value. The maximum entropy is realized for a uniform distribution on $X$, and its value is then $\log(n)$. Let us use a normalized entropy defined by
        \begin{equation}\label{eq::normalized_entropy}
            E^*(p) = \frac{E(p)}{\log(n)}
        \end{equation}
        The efficiency of a NMF/S$^{2}$NMF-based clustering can be assessed by looking into the diagonal blocks of the sorted matrix. We expect to observe less dispersion, hence less entropy, among the values a) taken by the features belonging to a cluster $i$, and b) restricted to those observations also belonging to cluster $i$. The rationale here is to group together observations having high scores on a given meta-feature (read in matrix $W$), hence having likely high scores on the features having a high loading in this meta-feature (read in matrix $H$). The  advantage of S$^{2}$NMF vs NMF is in the ability to group together not only the high values, but also the low values which can contain important information as well. \\
        
    
        "Clustering entropy" shown in Table \ref{table::nmf_vs_snmf_entropy_table} represents the relative entropy of the clustering induced by each model. It is calculated as follows:
        \begin{itemize}
            \item The normalized entropy of each diagonal cluster $(i, i)$ is calculated. The i-th diagonal cluster is a submatrix restricted to features in cluster, i and to observations having their dominant score on feature i. 
            \item The clustering entropy is the average of relative entropies measured on all \textbf{diagonal clusters}, weighted by their sizes.
            \item This average entropy on diagonal clusters can be compared with the normalized entropy of the whole matrix. The difference defines the "Entropy delta".
        \end{itemize}
        
        \begin{table}[htp!]
            \centering
            \begin{tabular}{|c|c|c|}
            \hline
            \textbf{} & \textbf{Entropy delta} \\ \hline
            \textbf{Restricted NMF} &  \textbf{-0.77\%} \\ \hline
            \textbf{Restricted S$^{2}$NMF} & \textbf{-8.03\%} \\ \hline
            \end{tabular}
            \caption{\label{table::nmf_vs_snmf_entropy_table} Relative entropies for NMF and S$^{2}$NMF restricted models.}
        \end{table}
        
        It is not a surprise to observe lower average entropies in the clusters, compared to the respective original matrices. This is an indirect proof of the efficiency of general NMF methods applied to this particular dataset. Moreover, we see that the loss in entropy is higher (-8.03\% vs -0.77\%) with S$^{2}$NMF compared to NMF, and this is a numerical confirmation of the visual difference between the respective heatmaps. 
        
        \subsubsection{Interpreting meta-features}\label{sssec::interpreting}

            Let us select as final model a 6-components S$^{2}$NMF, after elimination of the features from an initial run with sparsity 0.9. By taking a closer look at the definition of the meta-features, some coherence appears in the most influential features, i.e. those having the highest (positive) coefficients.

            \begin{table}[ht!]
            \centering
            \begin{tabular}{|c|c|c|c|}
            \hline
            \textbf{Rank}                                                       & \textbf{V1}                                                                    & \textbf{V2}            & \textbf{V3}                                                                                \\ \hline
            \textbf{0}                                                          & \textbf{hrts\_i-score}                                                         & \textbf{env1.1\_score} & \textbf{\begin{tabular}[c]{@{}c@{}}prs1.6\_employmt\_-\\ \_evolution\end{tabular}}         \\ \hline
            \textbf{1}                                                          & \textbf{hr\_i-score}                                                           & \textbf{env\_i-score}  & \textbf{\begin{tabular}[c]{@{}c@{}}bgt2.1\_tax\_paradis\_-\\ \_rejection\end{tabular}}     \\ \hline
            \textbf{2}                                                          & \textbf{hr\_l-score}                                                           & \textbf{cin\_i-score}  & \textbf{}                                                                                  \\ \hline
            \textbf{3}                                                          & \textbf{\begin{tabular}[c]{@{}c@{}}edh1.1\_score\_\\ non-discrim\end{tabular}} & \textbf{hrts\_l-score} & \textbf{}                                                                                  \\ \hline
            \textbf{4}                                                          & \textbf{cin\_i-score}                                                          & \textbf{env\_score}    & \textbf{}                                                                                  \\ \hline
            \textbf{\begin{tabular}[c]{@{}c@{}}Dominant \\ themes\end{tabular}} & \textbf{Human Rights}                                                          & \textbf{Environment}   & \textbf{\begin{tabular}[c]{@{}c@{}}Social Responsibility,\\ Tax Transparency\end{tabular}} \\ \hline
            \end{tabular}
            \end{table}
            
            \addtocounter{table}{-1}
            \begin{table}[ht!]
            \centering
            \begin{tabular}{|c|c|c|c|}
            \hline
            \textbf{Rank}                                                       & \textbf{V4}                                                             & \textbf{V5}                                                                & \textbf{V6}                                                                                \\ \hline
            \textbf{0}                                                          & \textbf{cg1.1\_score}                                                   & \textbf{cg1.1\_score}                                                      & \textbf{\begin{tabular}[c]{@{}c@{}}bgt2.1\_tax\_paradis\_\\ -\_rejection\end{tabular}}     \\ \hline
            \textbf{1}                                                          & \textbf{cg\_l-score}                                                    & \textbf{cg\_l-score}                                                       & \textbf{\begin{tabular}[c]{@{}c@{}}prs1.6\_employmt-\\ \_evolu\end{tabular}}               \\ \hline
            \textbf{2}                                                          & \textbf{cg\_i-score}                                                    & \textbf{cg4.1\_score}                                                      & \textbf{hr1.1\_score}                                                                      \\ \hline
            \textbf{3}                                                          & \textbf{cg\_score}                                                      & \textbf{env3.1\_score}                                                     & \textbf{}                                                                                  \\ \hline
            \textbf{4}                                                          & \textbf{cg2.1\_score}                                                   & \textbf{env2.2\_score}                                                     & \textbf{}                                                                                  \\ \hline
            \textbf{\begin{tabular}[c]{@{}c@{}}Dominant \\ themes\end{tabular}} & \textbf{\begin{tabular}[c]{@{}c@{}}Corporate\\ Governance\end{tabular}} & \textbf{\begin{tabular}[c]{@{}c@{}}Governance,\\ Environment\end{tabular}} & \textbf{\begin{tabular}[c]{@{}c@{}}Tax transparency,\\ Social responsibility\end{tabular}} \\ \hline
            \end{tabular}
            \caption{\label{table: meta-features definition}Assigning themes to meta-features based on their dominant features}
            \end{table}

            Table \ref{table: meta-features definition} show the 6 most influential features of each meta-feature. Interestingly, the topics of these influential features are interrelated. We can therefore assign a main theme to each meta-feature. For example, the theme related to feature V1 is mostly about Human Rights (hrts) and Human Resources (hr), while V2 is clearly concentrated on Environment (env).
            S$^{2}$NMF has traded dimension for interpretability, by drastically lowering the dimension (from 77 to 6) at the price of getting more fuzzy, but still providing meaningful features.\\
            
\begin{figure}[htp!]
               	\centering
                \includegraphics[width=\textwidth]{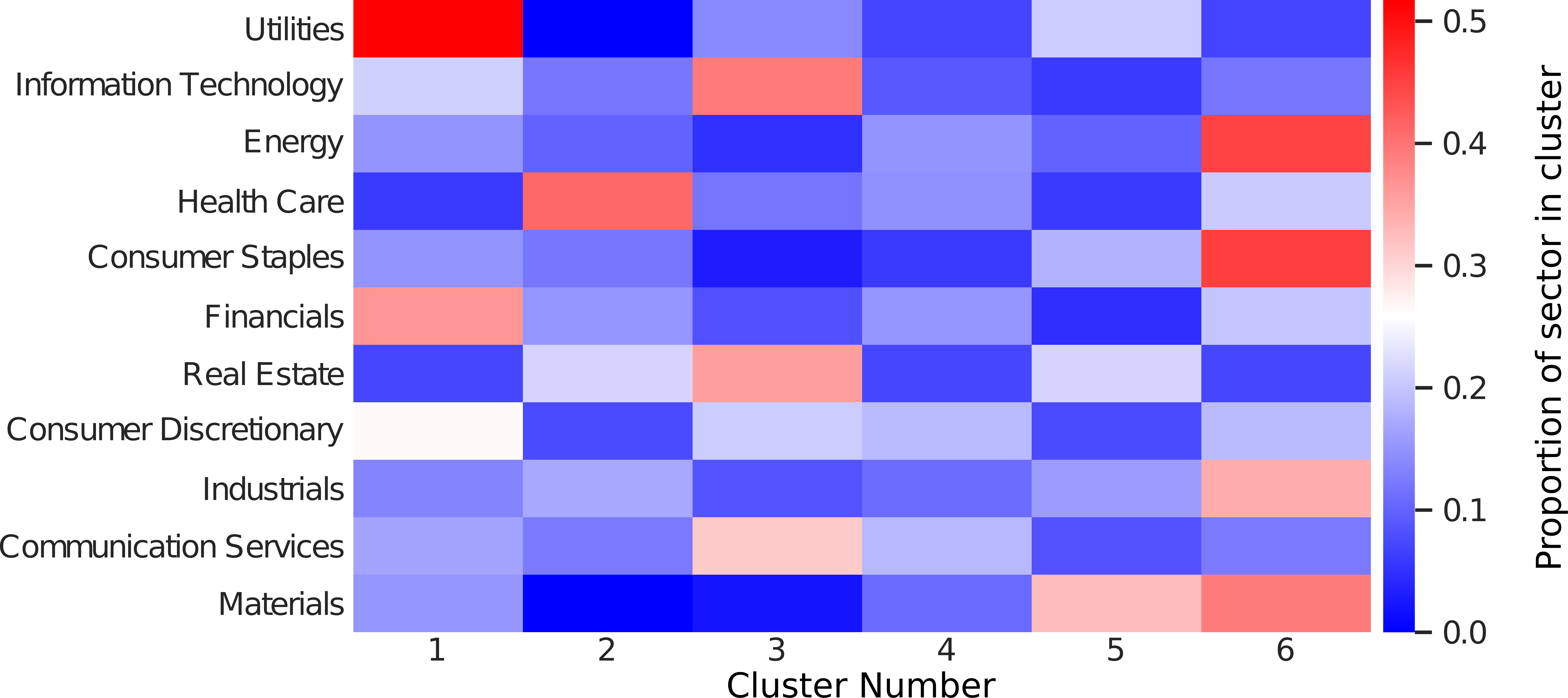}
            	\caption{\label{fig::snmf_sectors_vs_clusters} Incidence matrix between industrial sectors and S$^{2}$NMF clusters.}
            \end{figure}

            Chart \ref{fig::snmf_sectors_vs_clusters} shows the correspondence, or \textbf{incidence matrix}, between the 6 clusters identified by S$^{2}$NMF and the 11 industrial sectors of the stocks. Each cell shows the proportion of the stocks in a given sector (in rows) classified in a given cluster (in columns). It is visually clear that the two clusterings are not mutually redundant. S$^{2}$NMF provides a data-driven, endogenous way of grouping stocks with respect to their ESG scores. For example, we see that the Utilities sector has about half of its members classified on the cluster 1, whose dominant theme is Human Rights (table \ref{table: meta-features definition}). Similarly, the cluster 2 (Environment) is only dominant for the Health Care sector, and poorly represented in Utilities and Materials. 
            
\section{Conclusion and future work}\label{sec::conclusion}

    We have illustrated here the application of NMF to extra-financial score data. In contrast to principal component analysis, which focuses on risk reduction by creating mutually orthogonal factors, NMF allows the information contained in the data to be summarized faithfully using meta-variables, where the original variables appear as linear combinations with positive coefficients. These meta-variables are not constrained by mutual orthogonality;  they induce a natural clustering of the original variables and observations, at the cost of much lower error than would be produced by a PCA of the same dimension.

    Moreover, the S$^{2}$NMF variant presented here allows, by symmetrizing the role of low and high scores, to further improve the results in terms of error and cluster relevance.
    
    An induced result is the interpretability of the meta-features thus created, as shown by the comparison of their definitions with the prior knowledge of the meaning of the variables, which was not taken into account by the decomposition.
    
    Finally, we have shown that an initial application of NMF with a high level of sparsity allows for the removal of variables that are insufficiently relevant.\\
    
    Future directions of research are:
    
    \begin{itemize}
        \item The determination of heuristics independent of any prior knowledge of the domain, to optimize the number of components retained in the factorization. Possible heuristics are combinations of sparsity, robustness, precision and interpretability metrics.
        \item Generalizing the use of S$^{2}$NMF as a pre-processing of variables, in a features engineering step upstream of a supervised learning process. The aim here is twofold: to reduce computation time and, above all, to reduce the risk of overlearning. Replacing a large number of partially redundant features by a much smaller number of interpretable features containing globally the same amount of information is a good policy in machine learning, regardless of the domain.
        \item The clustering induced by S$^{2}$NMF, being independent from sectors, can induce a smart strategy for the imputation of missing data. 
        \item S$^{2}$NMF is applied to 2-dimensional data, and involves the use of 3-dimensional tensors to perform the factorization. A seemingly interesting generalization is to transpose respectively PosNegNMF and S$^{2}$NMF to 3-dimensional data (with the addition of a time axis). The same way S$^{2}$NMF uses an additional dimension to factorize the duplicated matrix (as described in equation \ref{eq::SNMF}), the 3-dimensional factorization would be achieved with 4-dimensional tensors, and yield PosNegNTF and SNTF.
    \end{itemize}
    
    We thank Vigeo-Eiris for making some datasets available to the authors.
    We also thank Vincent Margot for useful discussions and remarks.

\bibliographystyle{plainnat}
\bibliography{biblio} 

\end{document}

%% file: config.tex
\usepackage{authblk}
\usepackage{DAMIPaperStyle}
\usepackage{makecell}
\usepackage{url}

\usepackage{ulem}
\usepackage{xparse}
\usepackage[left=3.5cm, width=16.5cm, height=25.4cm, top=1.3cm]{geometry}
\usepackage{mathrsfs}
\usepackage{placeins}
\usepackage{algorithm}
\usepackage{algorithmic}
\usepackage{graphicx}
\usepackage{mathtools}
\usepackage{commath}
\usepackage{dsfont}
\usepackage{wrapfig}
\usepackage{hyperref}
\usepackage[figuresleft]{rotating}
\DeclareDocumentCommand\CG{mg}{\textcolor{blue}{\IfNoValueTF{#1}{}{\uwave{#1}}\IfNoValueTF{#2}{}{\footnote{\textcolor{blue}{#2}}}}}
\DeclareDocumentCommand\NM{mg}{\textcolor{red}{\IfNoValueTF{#1}{}{\uwave{#1}}\IfNoValueTF{#2}{}{\footnote{\textcolor{red}{#2}}}}}
\DeclareMathAlphabet{\pazocal}{OMS}{zplm}{m}{n}

\def\X{\ensuremath{\mathbf X}} 
\def\M{\ensuremath{\mathbf M}} 

\providecommand{\keywords}[1]{\textbf{\textit{Keywords: }} #1}

\newcommand*{\tran}{^{\mkern-1.5mu\mathsf{T}}}